\documentclass{PoS}

\title{Cooling Tests of the NectarCAM camera for the Cherenkov Telescope Array}

\ShortTitle{NectarCAM thermal demonstrator}

\author{\speaker{E. Moulin$^{1}$} , 
C. Diaz$^2$, D. Durand$^3$, O. Feirreira$^2$, M. Fesquet$^3$, B. Giebels$^4$, J.-F. Glicenstein$^1$, 
D. Loiseau$^5$, F. Louis$^3$, F. Nunio$^5$, S. Rateau$^4$,
        for the NectarCAM and CTA consortia$^6$\\\\
        E-mail: \email{emmanuel.moulin@cea.fr}\\
        $^1$ DSM/Irfu/SPP, CEA Saclay, F-91191 Gif-Sur-Yvette Cedex, France\\
        $^2$ CIEMAT, 28040 Madrid, Spain\\
        $^3$ DSM/Irfu/SEDI, CEA Saclay, F-91191 Gif-Sur-Yvette Cedex, France\\
        $^4$ IN2P3/LLR, Ecole Polytechnique, 91128 Palaiseau, France\\
        $^5$ DSM/Irfu/SIS, CEA Saclay, F-91191 Gif-Sur-Yvette Cedex, France\\
        $^6$ Full consortium author list at http://cta-observatory.org

}

\abstract{The NectarCAM is a camera proposed for the medium-sized telescopes in the
framework of the Cherenkov Telescope Array (CTA), the next-generation observatory
for very-high-energy gamma-ray astronomy. The cameras are designed to operate in an open
environment and their mechanics must provide protection for all their components under the
conditions defined for the CTA observatory. In order to operate in a stable environment and
ensure the best physics performance, each NectarCAM will be enclosed in a slightly overpressurized,
nearly air-tight, camera body, to prevent dust and water from entering. The total
power dissipation will be $\sim$7.7 kW for a 1855-pixel camera. The largest fraction is dissipated
by the readout electronics in the modules. We present the design and implementation of the
cooling system together with the test bench results obtained on the NectarCAM thermal
demonstrator.}

\FullConference{The 34th International Cosmic Ray Conference,\\
		30 July- 6 August, 2015\\
		The Hague, The Netherlands}

\begin{document}

\section{Introduction}
The Cherenkov Telescope Array (CTA) is the next-generation observatory for ground-based very-high-energy
gamma-ray (E $\gtrsim$ 100 GeV) astronomy. It will have about a factor-ten improved sensitivity compared to current
 instruments and enlarged accessible energy range from a few tens of GeV to a few hundred of TeV~\cite{2011ExA32193A}. The observatory will consist of several tens of telescopes of different sizes on two sites.  Each site will have some combination of 23-m dish 
Large-Sized Telescopes (LSTs), 12-m dish Medium-Sized Telescopes (MSTs), $\sim$10-m Dual-Mirror Medium-Sized Telescopes (SCTs) and $\sim$4-m dish Small-Sized Telescopes (SSTs). One array is envisaged in each hemisphere  to allow for full coverage of the sky. The NectarCAM cameras are designed to equip some of the MSTs of CTA.
The cameras are operating in an open environment and the mechanics must provide protection to all the
components under the conditions defined for the CTA observatory. The two main requirements for the
camera mechanics related to the cooling system are the following : {\it (i)} the cameras host the photodetectors and readout electronics, which requires that the camera mechanics prevent dust and water from entering inside, and thus needs to be air tight, and {\it (ii)} the surface temperature of electronics hosted in the camera must be kept constant during data taking to ensure the best physics performance.

The NectarCAM camera (dimension 2.9$\times$2.9 $\times$1.5 m$^3$) is divided into two parts. The front part hosts the readout electronics and photodetectors, and the rear part hosts the backplanes, cabling, auxiliary electronics
and power supplies. The instrumented area lies behind a protective entrance window in the front part of the
camera, which also ensures the air tightness of the camera volume. The shape of the instrumented area is
hexagonal, with a maximum diameter of 2.25 m. Behind the protective surface, light guides coupled to the
photodetectors define the pixels of the camera. Sets of seven of these photodetectors are mechanically and
electrically coupled with the readout electronics, in an ensemble referred to as a cluster (see Ref.~\cite{nectarcam} for more details). The NectarCAM camera body will be nearly air-tight and slightly over-pressurized to further prevent dust and water from entering. The total power dissipation will be $\sim$7.7 kW for a 1855-pixel camera. 
The largest fraction of thermal power will be dissipated by the readout electronics in the modules. The
heat produced by the photon detector plane equipped with photomultiplier tubes is negligible and does
not require specific cooling. The current design foresees cooling with fans and a cooled air flow provided by water-air heat exchangers. The main MST camera specification for the cooling system relevant to the results presented here is related to temperature homogeneity.  The maximum air temperature difference between any two points of the
module holder part should be less than 10$^{\circ}$C. 

The paper presents a first series of measurements and results to assess the performance of the cooling
system on a thermal demonstrator relative to temperature gradients in a module holder 
and give quantitative estimates of the spatial temperature homogeneity in operating mode. 
Test bench results in a climatic chamber are also shown to study the influence of external temperature conditions 
during data taking.

\section{Design and set-up of the thermal demonstrator}
\begin{figure}[b]
\centering
\includegraphics[width=9cm,angle=270]{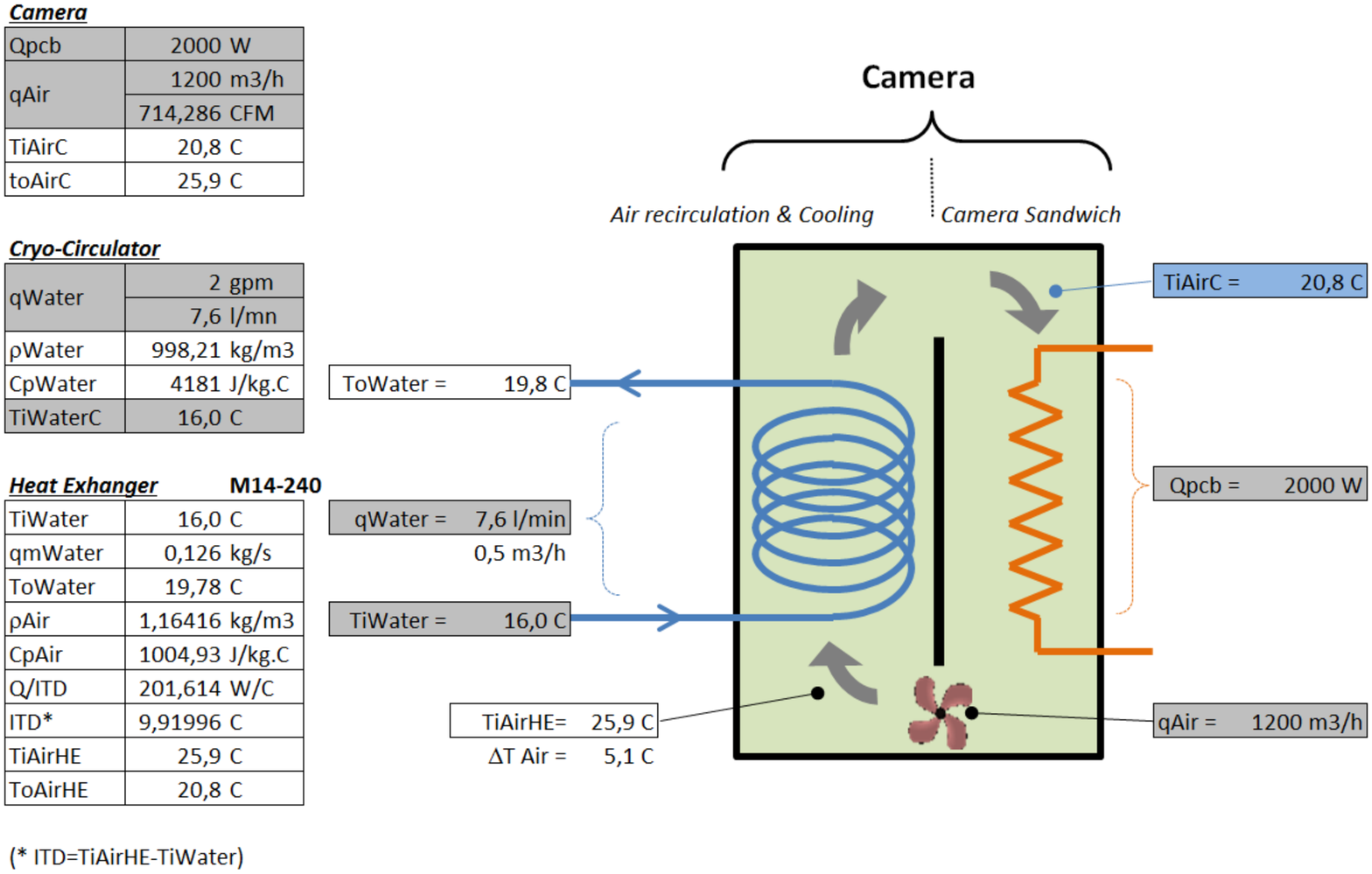}
\caption{The thermal model of the NectarCAM MST cooling system. The sizing work flow, heat exchanger(s) and chiller characteristics are shown.} 
\label{fig:prototype}
\end{figure}

\begin{figure}[t]
\centering
\includegraphics[width=7cm,angle=0]{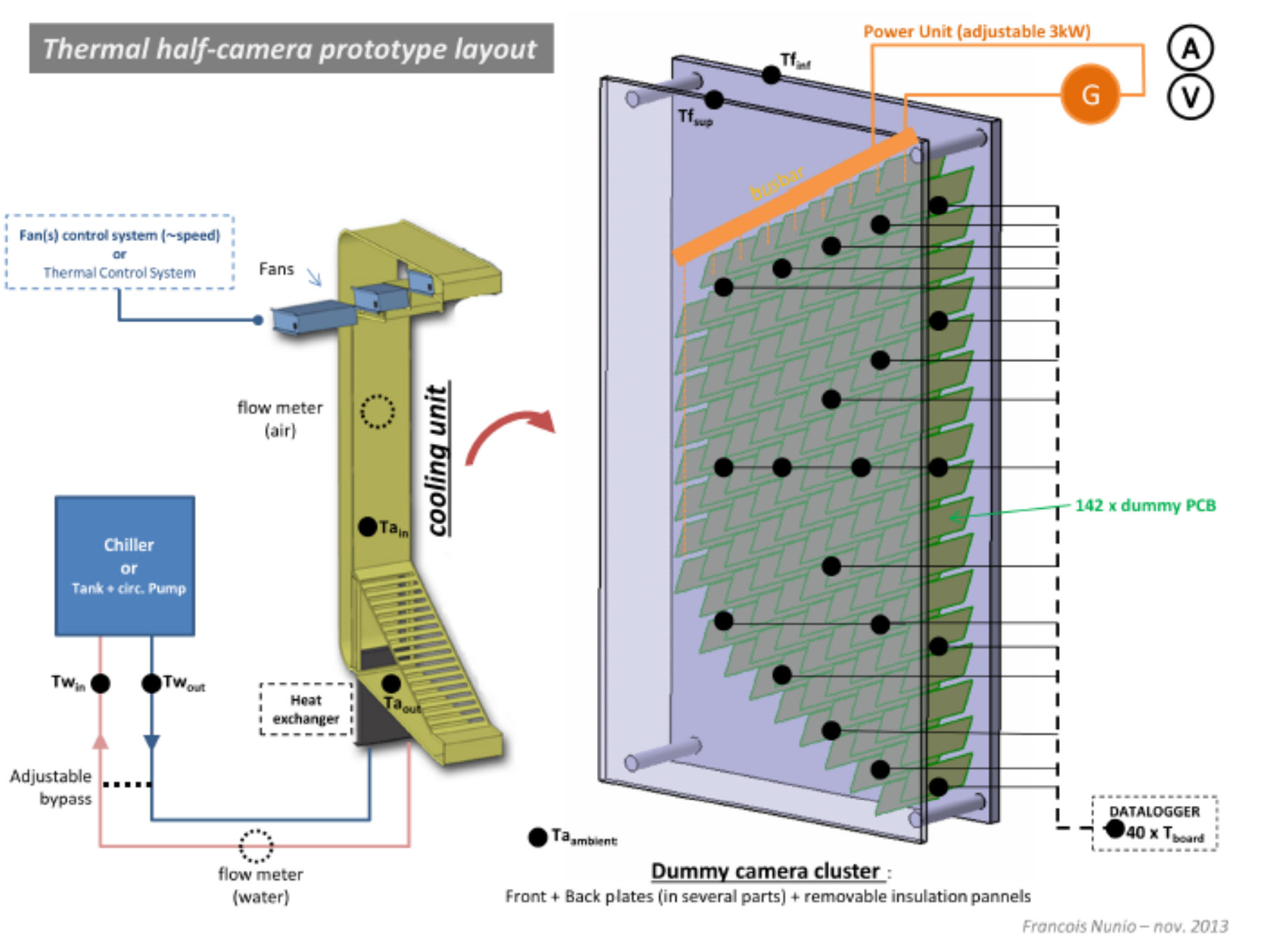}
\caption{The layout of the NectarCAM half camera thermal prototype with chiller and cooling unit. The dummy camera module holder is made of a front and back plates. Runners are made in the rear holder structure to insert the dummy FEBs.} 
\label{fig:layout}
\end{figure}

\begin{figure}[t]
\centering
\includegraphics[width=7cm,angle=360]{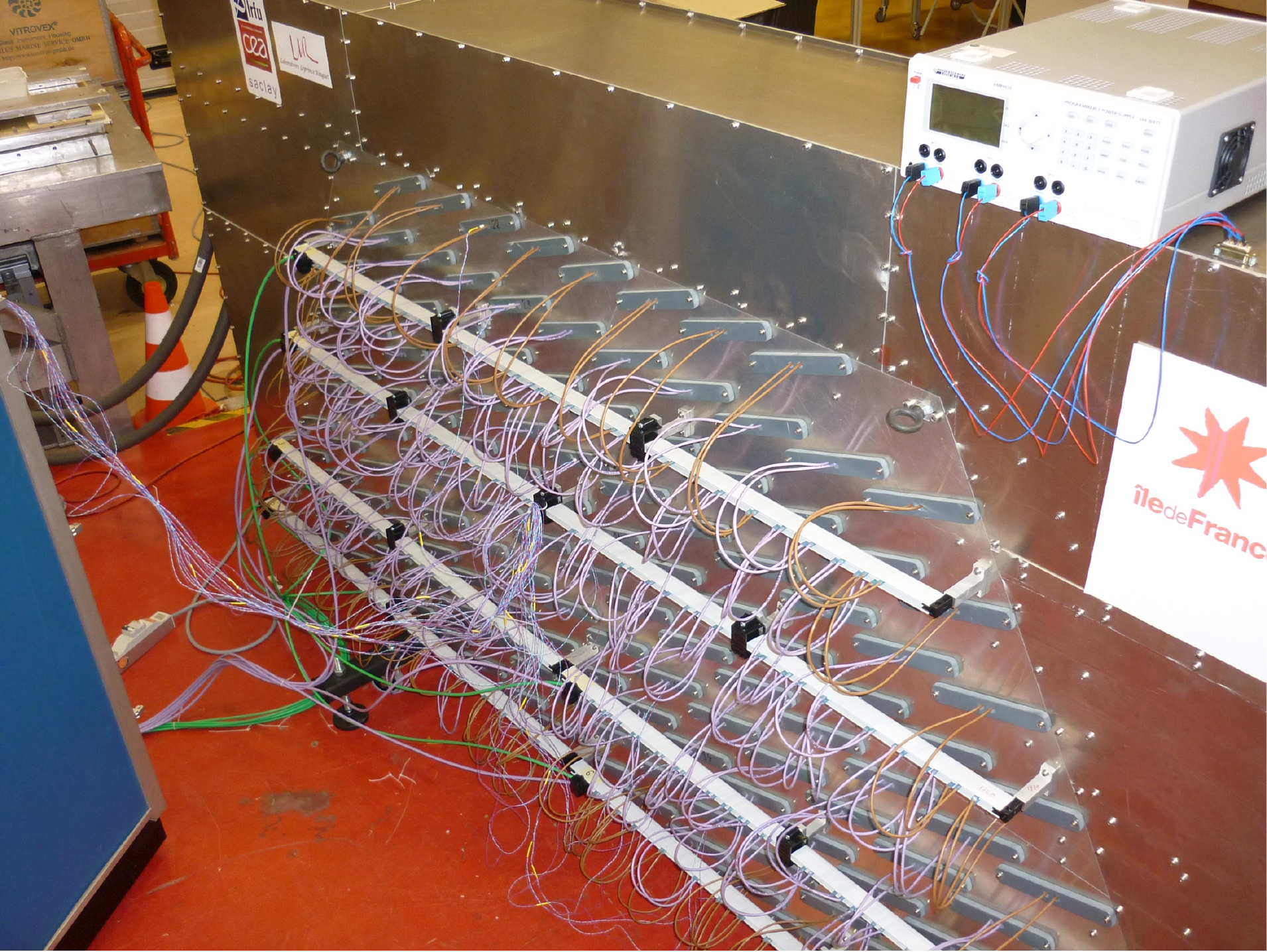}
\caption{A picture of the thermal demonstrator set-up. All the dummy FEBs  are supplied by the same power unit. } 
\label{fig:photo}
\end{figure}

The thermal demonstrator of the NectarCAM camera consists in a cooling system composed of a cooling unit and a chiller, and a half-camera module holder equipped with dummy electronic front-end boards (FEB).  
The sizing of the thermal components is conducted considering the energy balance
between the coolant fluid and the circulating air. The coolant parameters are the fluid (water) 
air-flow rate and the heat exchanger inlet temperature. The warming parameters are the heating power 
of the readout electronics in the crates and the air-flow rate in the half-camera module holder.
Figure~\ref{fig:prototype} represents the sizing work flow, which is used to check the overall agreement between the fan(s), heat exchanger(s) and chiller characteristics. The water cooling loop of the test bench is controlled by a JULABO FE2800 recirculating cooler. The air loop is driven by a set of three centrifugal fanning modules. Each of the three
fans gives a nominal flow rate of 700 m$^3$h$^{-1}$, at a nominal speed of 2865 rpm. Fan speeds can be controlled
independently by connecting the dedicated input to an adjustable voltage source (0 -10 V).
The copper tube fin liquid to air heat exchanger is a LYTRON M14-240. The power supply for the demonstrator is
controlled with Ethernet interface board and able to deliver a power of 3 kW with a maximum current of 100 A and a maximum voltage of 80 V. 
The dummy half camera consists of an assembly of front and back plate
sandwiches referred as to  the module holder, with slits to insert and fix dummy FEBs from the rear. Fig.~\ref{fig:layout}
shows a schematic of the thermal demonstrator with the chiller, the cooling unit with fanning modules, and the module holder. Fig.~\ref{fig:photo} shows a picture of the thermal demonstrator for NectarCAM cameras.

The dummy FEBs were designed to have the same integrated heating power as the real FEBs which
is $\sim$15 W per board. 
It uses the resistance of a single long copper trace to dissipate the requested thermal power inside the NectarCAM module holder structure. The dummy FEBs were produced with single-sided standard FR4 PCB 1.6 mm thick. 
The PCB board length and width are 350 and 136 mm, respectively. The copper thickness is 170 $\mu$m.
The copper trace length and thickness were chosen based on the resulting total resistance and the current,
voltage and power limits of the available power source. The dummy FEB resistance is $\sim$59~$\Omega$.
PT100 
sensors are glued on the top side of 39 dummy FEBs. 
An autonomous data logger provides 40 analog measurement channels. It is equipped
with an internal memory to allow direct capture of data. 

\section{Test bench results}
\begin{figure}[t]
\centering
\includegraphics[width=6cm,angle=0]{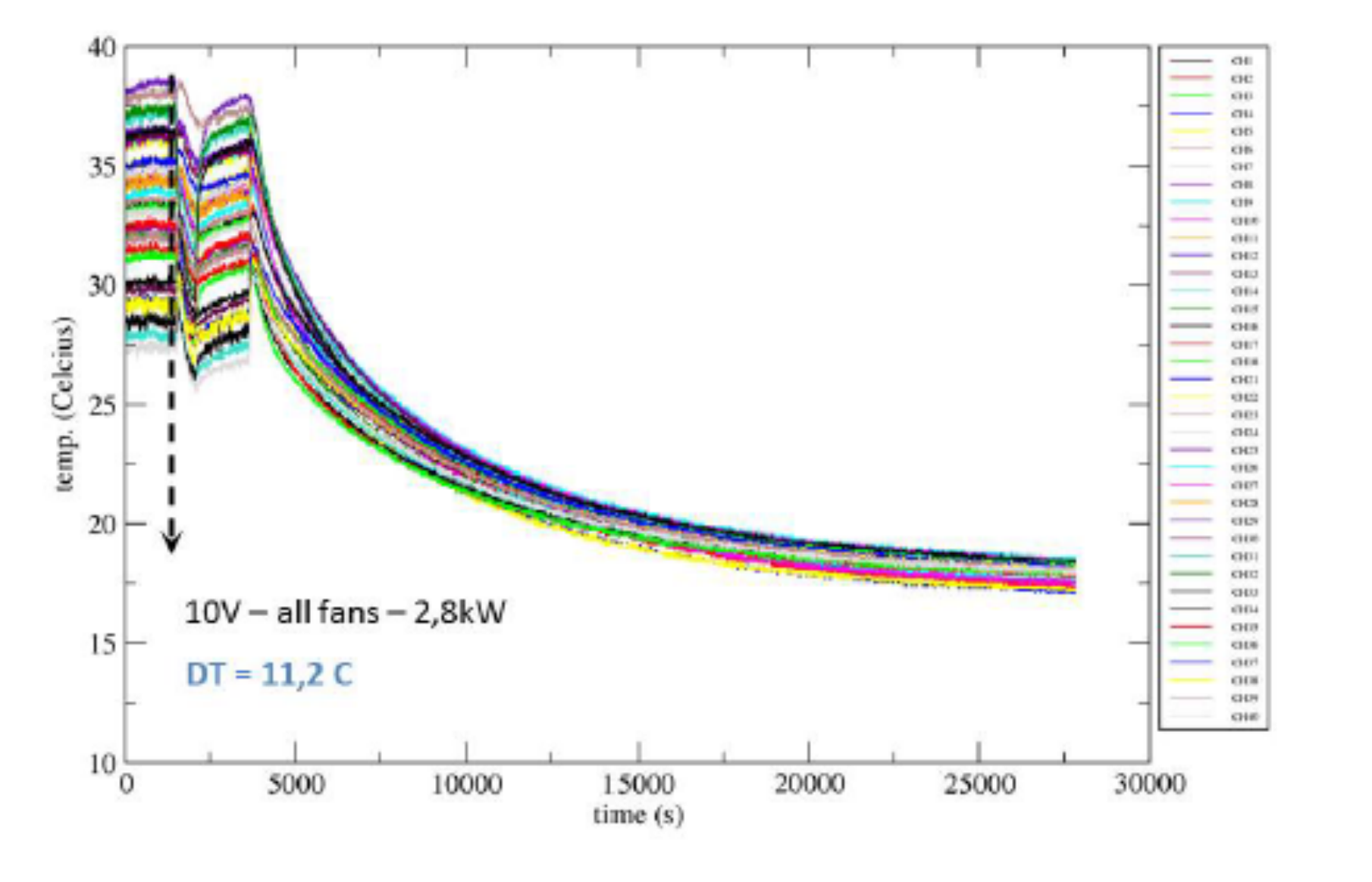}
\hspace{1cm}\includegraphics[width=8cm,height=4cm,angle=0]{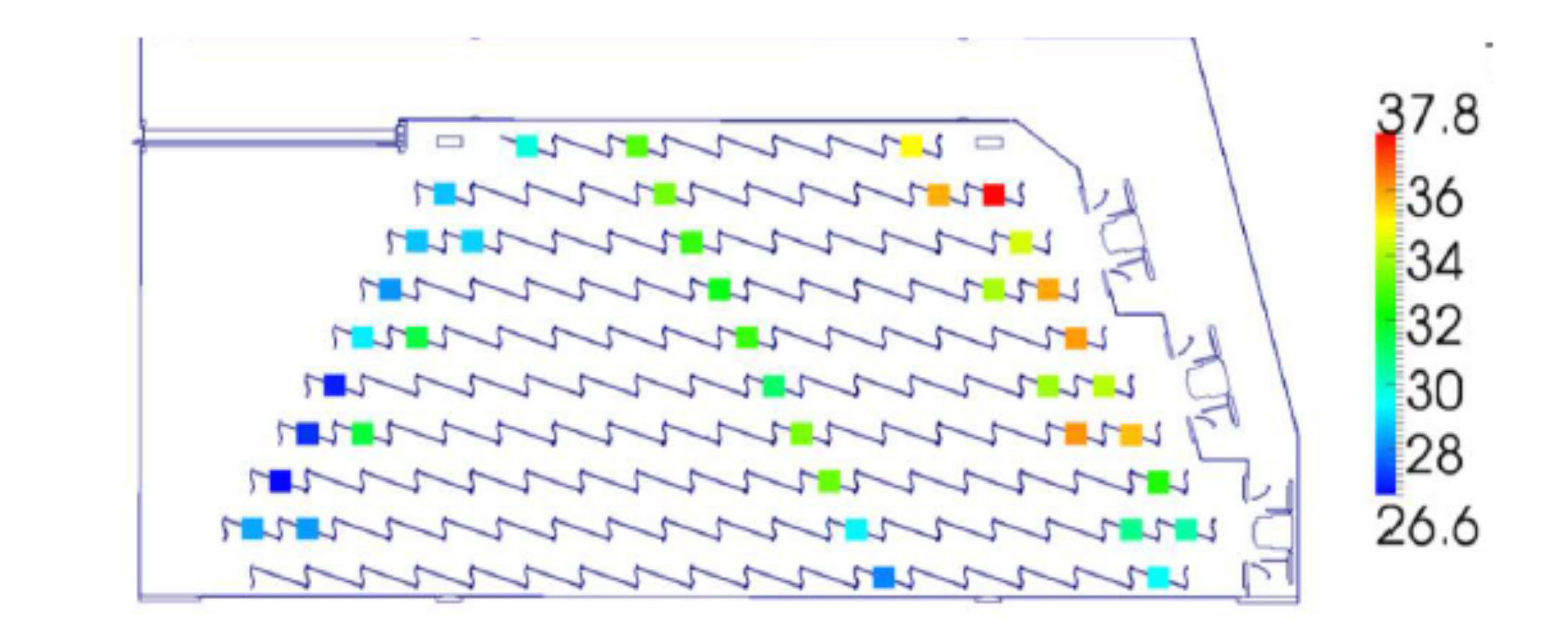}
\caption{{\it Left:} An example of a data acquisition run. The temperature measurements ($^{\circ}$C) with the thermal demonstrator are done with a total injected power of 2.8 kW (about 40\% higher than the specifications). The power to the dummy FEBs is turned off at a time of $\sim$3700 s, and all fans are running at running at full speed.  {\it Right:} Temperature map of the run. The mapping of the temperature is done at a time of $\sim$1200 s. The temperature gradient (DT) is 11.2$^{\circ}$C.} 
\label{fig:TemperatureMap}
\end{figure}

\begin{figure}[b]
\centering
\includegraphics[width=7cm,angle=0]{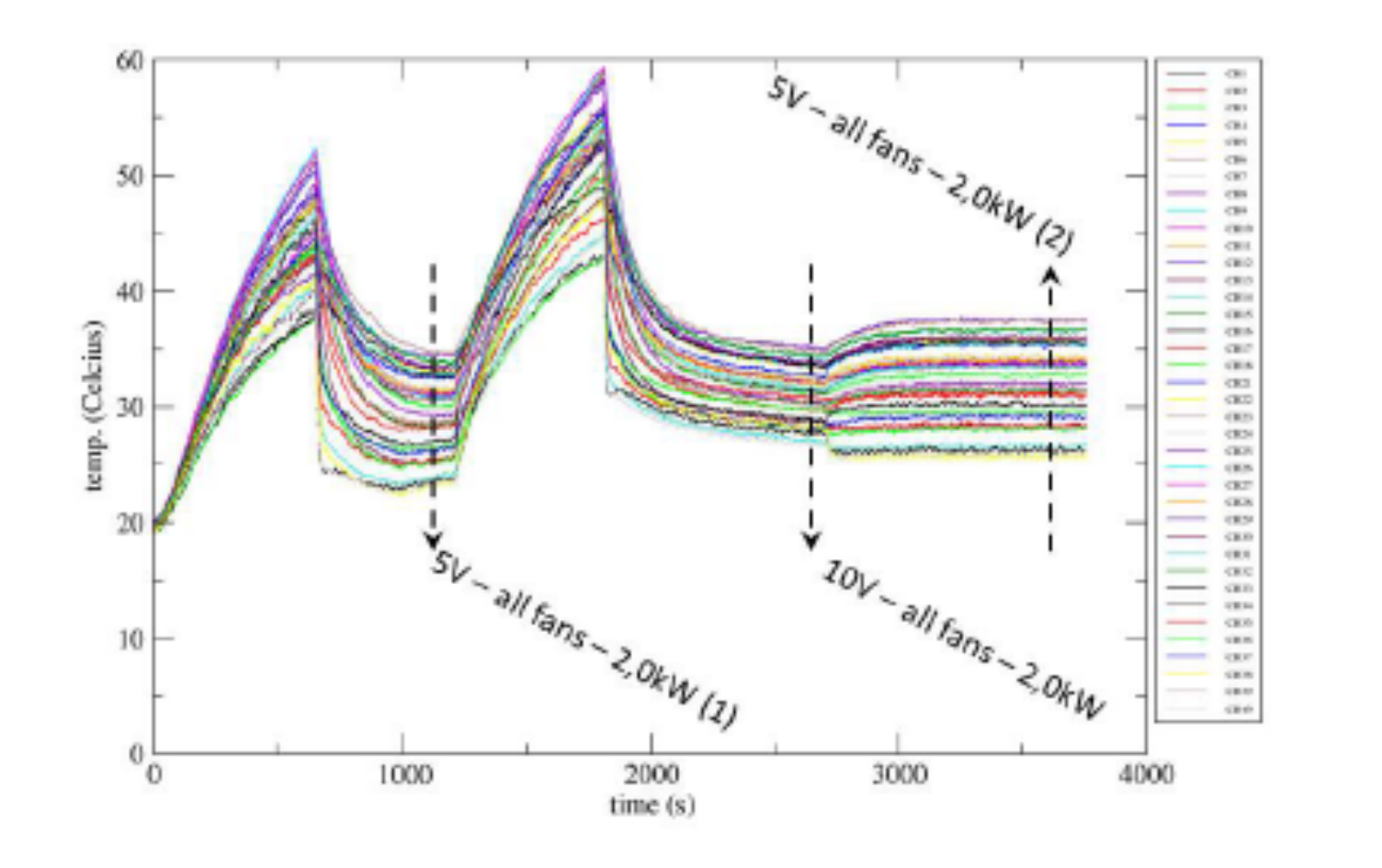}
\hspace{1cm} \includegraphics[width=6cm,height=6cm,angle=0]{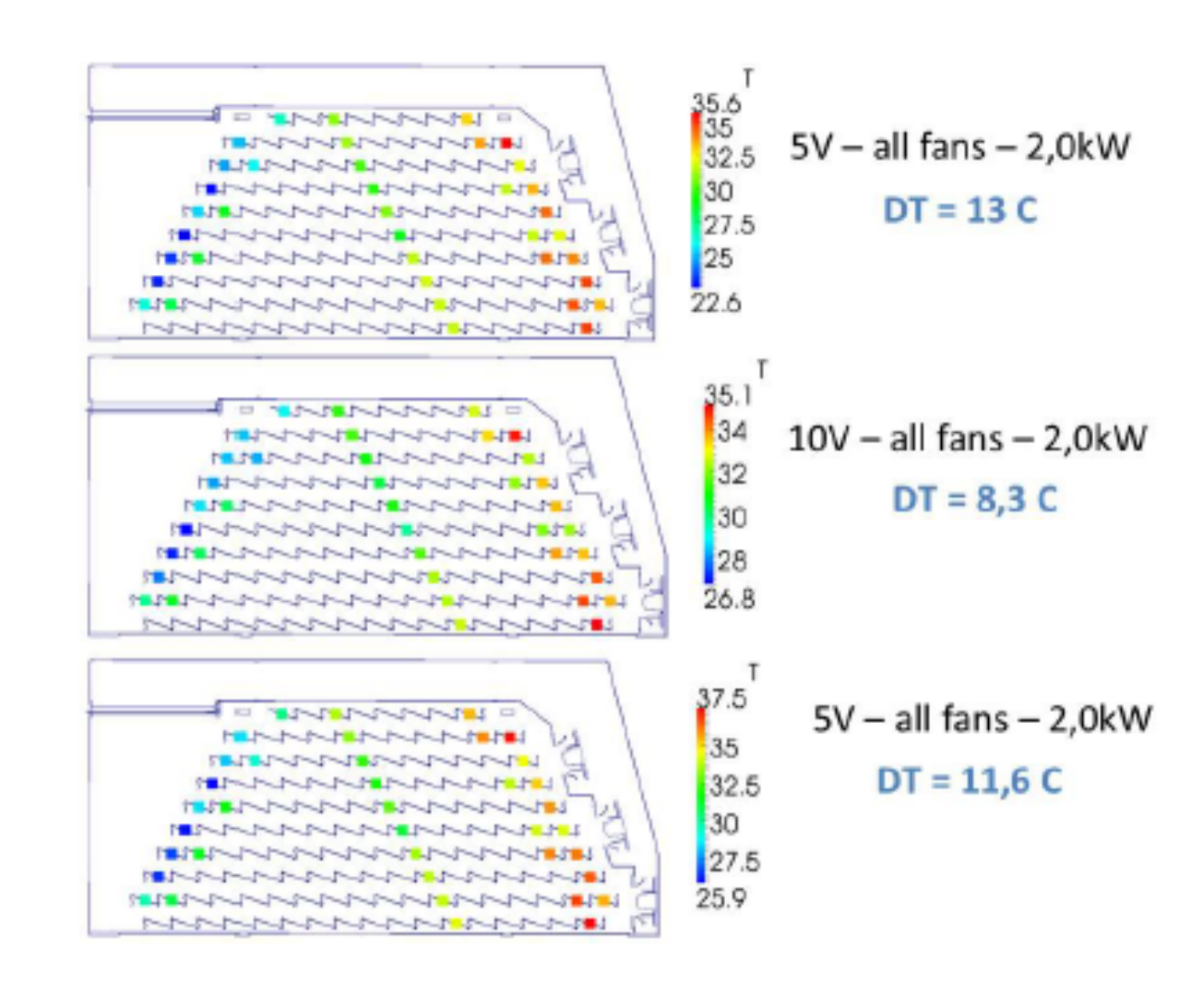}
\caption{The data acquisition run with the power injections at 0 and $\sim$1200 s, respectively,  from which static temperature measurements are extracted. The two peaks correspond to the times when the fans are started up.
{\it Left:} The downward arrows show the time when temperature measurements are done and specify the operating conditions: power injections of 2 kW and the speed of the fans in Volts (V). {\it Right:} Temperature maps from the measurements done under the three conditions specified in the left panel.} 
\label{fig:TemperatureMapFanSpeed}
\end{figure}

A first series of tests with the completely integrated set-up were conducted in an open unregulated environment.
During these tests, the experimental hall temperature was around 20$^{\circ}$C with very slow variations of several degrees over a 24-hour period. The chiller temperature was set to 18.5$^{\circ}$C. 
The left panel of Fig.~\ref{fig:TemperatureMap} shows an example of an acquisition run where the total injected power is 2.8 kW (which is about 40\% above specifications) and the three fans running at full speed. The recorded temperature map is shown is the right panel of Fig.~\ref{fig:TemperatureMap} obtained from the temperature sensors placed on the dummy FEBs. The air temperature increases as it flows from the heat exchanger to the fans. The temperatures were measured at a time of $\sim$3700 s and the temperature gradient was
$\sim$11$^{\circ}$C. Lowering the total injected power down to 2 kW with the three fans running at full speed reduces the temperature gradient (right panel of Fig~\ref{fig:TemperatureMapFanSpeed}). The impact of the fan speed  
on the temperature gradient is presented on Fig.~\ref{fig:TemperatureMapFanSpeed}.  Reducing the speed increases the temperature dispersion. Due to the slow dynamics of the system,  reproducible measurements require stabilized conditions. For a total injection power of 2 kW and the selected fans running at full speed, the specifications are met with the current design of the thermal demonstrator.

\begin{figure}[t]
\centering
\includegraphics[width=16cm,angle=0]{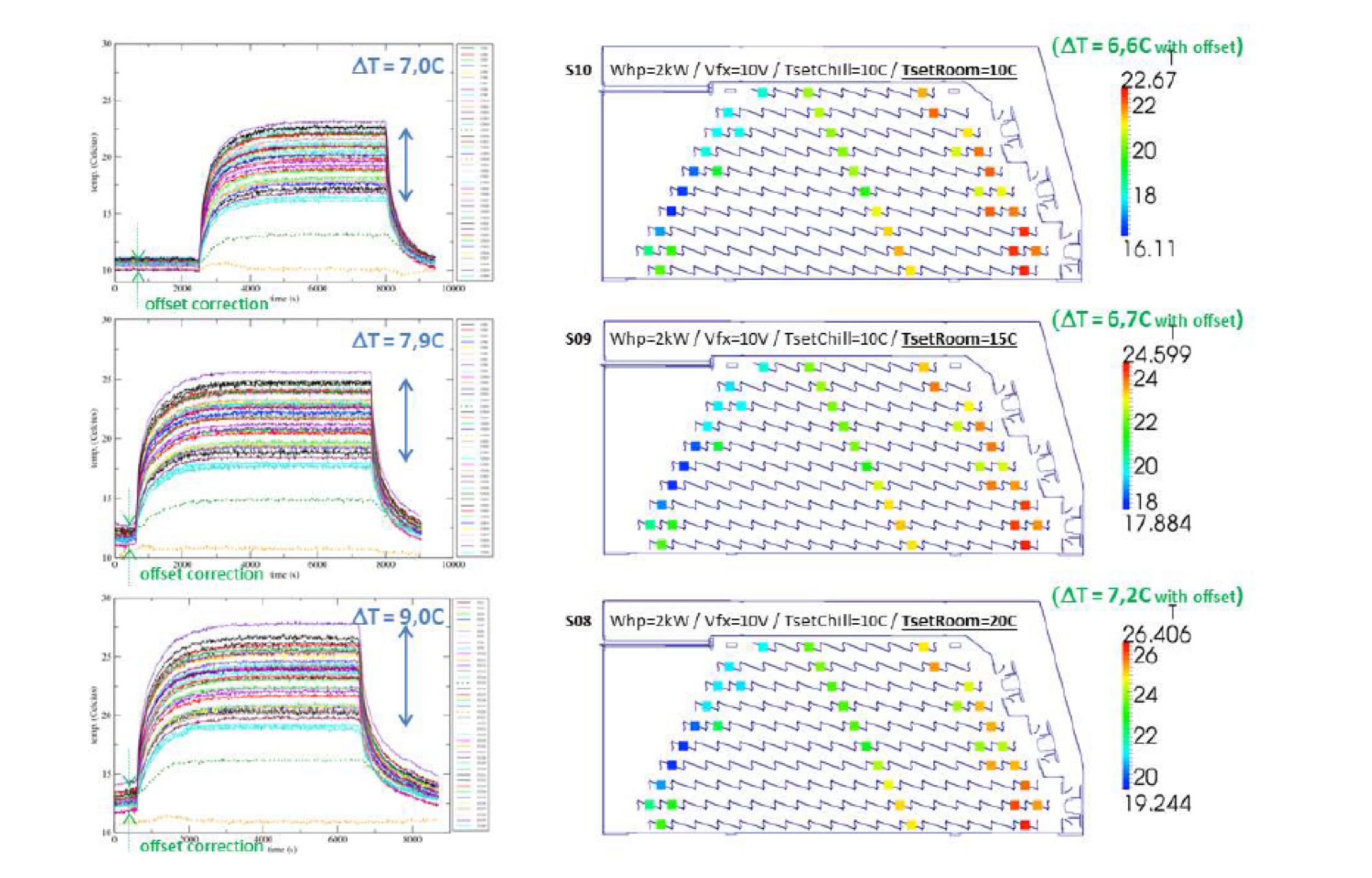}
\caption{The stabilized temperature measurements with dummy FEBs dissipating 2 kW and climatic chamber
temperature set to 10$^{\circ}$, 15$^{\circ}$C, and 20$^{\circ}$C, respectively. 
The temperature measurements are done $\sim$60 min after the dummy FEBs are powered.
The temperature gradients are given in blue on the left hand side. The temperature dispersions after 
offset correction are given on the right hand side. Note that, on the
left hand plots, the isolated yellow and green traces are from PT100 temperature sensors mounted on the
chiller's water inlet and outlet.
} 
\label{fig:TemperatureMapClimaticChamber}
\end{figure}
These tests revealed that the performance of the available water cooler depends on the outside air temperature. In order to recover the chiller's full efficiency, the NectarCAM thermal demonstrator was moved in a thermally-regulated climatic chamber. This allows it to reach a lower ambient air temperature compared to the case in an open field, which is closer to the expected observational conditions. 
Several sensors were added to the system including two surface temperature sensors
(PT100 probes + plungers) mounted on the water pipe inlet and outlet, and a water flow meter. An additional sensor was used to record the ambient temperature inside the climatic chamber.
Several tests were performed in order to investigate the conditions for a minimal temperature gradient over
the FEB sandwich. 
Stabilized temperature measurements were done with the dummy FEBs dissipating 2 kW. A series of tests was conducted with several ambient air temperatures in the climatic chamber.
Fig.~\ref{fig:TemperatureMapClimaticChamber} shows the data acquisition runs for the ambient temperature 
temperatures set to 10$^{\circ}$, 15$^{\circ}$C, and 20$^{\circ}$C, respectively,  together with the corresponding temperature maps. On the left panel are given the temperature gradients
$\Delta$T.  The temperature monitoring from the  PT100 sensors mounted on the chiller's water inlet (yellow curve) and outlet (green curve) are also shown. respectively. On the right panel, the temperature maps after offset correction\footnote{The set of PT100 sensors, although placed in a conditions such that they should provide identical measurements, have an intrinsic dispersion that needs to be accounted for when assessing the temperature gradient 
in the module holder.} are given.  For this series of tests, the temperature gradient is well below the 10$^{\circ}$C specifications. 

Further measurements were performed to measure the air flow in the module holder.
The measurements were made on the busbar side of the thermal demonstrator using a specific probe holder.
 In order to control the air flow balance over the FEBs, the chiller and dummy front end boards were powered off during these measurements. The mean air flow measured through the rectangular section nearest to the heat exchanger is 
 1858~m$^3$h$^{-1}$ which corresponds to an average air velocity of 7.07 m s$^{-1}$. This is in good agreement with the nominal air flow generated by the three fans. 
The velocities in the nine channels formed by the FEBs was measured and it was checked that the air flow is balanced over the channels. The velocity measurement is done at one point inside a given channel and it has been checked that it gives a realistic estimate of the effective air flow through that channel. 
The results show that the air flow is between 11\% and 12\% of the total flow in the different channels and well balanced among them,  except for the first two upper channels which receive a lower than expected fraction of the total air flow despite their reduced length. The results suggest that possible improvements of the set-up aiming at reducing the temperature gradient include the increase of the flow through the last three lower channels by increasing the speed of the lower fan relative to the other two, and the addition of a deflector in the lower corner of the air duct.

\section{Summary and outlook}
A thermal demonstrator  has been designed to test the validity of the cooling system for the MST NectarCAM cameras proposed for CTA. The test bench results obtained with the present prototype allow the current option 
for the cooling system of the NectarCAM camera to meet the CTA specifications. The temperature gradient between any two points of the module holder is lower than the maximal value specified at 10$^{\circ}$C.

There are several ways currently under investigation to further decrease the temperature gradient 
over the module holder. Among them are enhancing the air flow balance through the different channels either using static
or adjustable mechanical detectors, or using a finer control of the fans speeds. Further tests are foreseen
with an upgraded chiller, additional sensors (e.g. air inlet and outlet temperature sensors by the heat exchanger, temperature sensors bonded to the module holder walls), as well as with modifications of the prototype to closely match the final camera structure. If a general heat balance of the demonstrator is required, the wall of the demonstrator would benefit from insulation to reduce heat transfer through the conductive aluminum walls of the module holder.

\section*{Acknowledgements}
We gratefully acknowledge support from the agencies and organisations listed in this page:
http://www.cta-observatory.org/?q=node/22. This work was supported by DIM ACAV grants from R\'egion \^Ile-de-France.

\end{document}